%% file: qapl10.tex
\documentclass[copyright,creativecommons]{eptcs}
\providecommand{\event}{QAPL 2010} 
\providecommand{\volume}{}
\providecommand{\anno}{}
\providecommand{\firstpage}{1}
\providecommand{\eid}{}
\usepackage{breakurl}        
\usepackage{amsfonts}

\input{Environments}

\input{Commands}

\title{Approximate Testing Equivalence Based on \\ Time, Probability, and Observed Behavior}
\author{Alessandro Aldini
\institute{Dipartimento di Matematica, Fisica e Informatica -- University of Urbino, Italy}
\email{aldini@sti.uniurb.it}
}
\def\titlerunning{Approximate Testing Equivalence Based on Time, Probability, and Observed Behavior}
\def\authorrunning{A.~Aldini}

\begin{document}
\maketitle

\begin{abstract}

Several application domains require formal but flexible approaches to the comparison problem.
Different process models that cannot be related by behavioral equivalences should be compared 
via a quantitative notion of similarity, which is usually achieved through approximation of 
some equivalence. While in the literature the classical equivalence subject to approximation 
is bisimulation, in this paper we propose a novel approach based on testing equivalence.
As a step towards flexibility and usability, we study different relaxations taking into account 
orthogonal aspects of the process observations: execution time, event probability, and observed 
behavior. In this unifying framework, both interpretation of the measures and decidability of 
the verification algorithms are discussed.

\end{abstract}

%
\section{Introduction}
%

The need for a comparison between process models is an important requirement in several practical domains, ranging 
from the model-based verification of web service composition~\cite{FUMK} to security~\cite{BPW}, safety~\cite{SWD}, 
and performability~\cite{AABBBL} verification. For instance, equivalence checking can be helpful to compare a web 
service implementation with some desired qualitative/quantitative service description, to relate an implemented 
software architecture to a reference dependable architectural model, and to reveal the performability impact of one 
component over the whole system through the comparison of the two system views that are obtained by 
activating/deactivating the component (this is generally called noninterference analysis). Such a comparison must 
be based on a precise semantics and some notion of process equivalence. 
In the formal methods community several notions of equivalence have been proposed which differ from each other for 
their observational power -- e.g.\ from the ``weakest'' trace equivalence to bisimulation through the 
``intermediate'' testing equivalence -- and for their granularity -- e.g.\ from nondeterministic versions of 
observation equivalences to the corresponding probabilistic, real-time, and stochastically timed extensions (see, 
e.g., \cite{HPA} for a survey in the setting of process algebra).

In real-world applications perfect equivalence is usually hard to achieve when comparing models that describe 
either a system at different abstraction levels or alternative implementations of the same ideal system. 
Hence, adding the quantitative aspect to the comparison is of paramount importance to establish how much these 
models fit according to an expected behavior. This can be done, e.g., in a framework where fine-grain models 
specify probability distributions of events and/or their temporal behaviors. Alternatively, functional models can 
be quantitatively compared with respect to a benchmark of testing scenarios. In any case, some kind of mathematical 
function is employed to estimate the degree of similarity between models that do not exhibit the same behavior.

In this paper, a new approach to the approximation of behavioral equivalences is proposed in a process-algebraic 
setting in which three alternative dimensions -- time, probability, and observed behavior -- characterize what 
we mean by degree of similarity from the viewpoint of the expressive power of an external observer. 

First, we compare process models on the basis of their temporal behavior. Taking into account the passage of time 
when observing the process execution requires the specification of durations. In our setting, system activities are 
associated directly with their durations, which are modeled through exponentially distributed random variables. 
In particular, the stochastic process governing the system evolution over time turns out to be a continuous-time 
Markov chain (CTMC). 
Second, by considering that timing aspects are dealt with probabilistically, it makes sense to compare process models 
with respect to probability distributions associated with their behaviors. 
Third, in order to analyze the observed behavior by abstracting from additional quantitative information such as
time and probability, we introduce an approach that allows the distance between process models to be estimated with 
respect to their functional reaction to test-driven executions.

All the three dimensions are considered in the setting of a unifying semantics. More precisely, we employ a Markovian 
extension of testing equivalence~\cite{Ber07}, whose use represents a novelty in the field of approximate analysis. 
The main reason for this choice is that Markovian testing equivalence provides in a natural and explicit way an ideal 
framework for the definition of degree of similarity with respect to time, probability, and observed behavior. 
To give some intuitive insights, Markovian testing equivalence compares processes in terms of probability of observing
test-driven computations that somehow ``pass'' tests and satisfy temporal constraints about the amount of time needed to
pass these tests. Therefore, by relaxing in turn each of these parameters -- durations associated with specific computations, 
probability distributions of these computations, and kind of tests elucidating them -- we easily obtain different notions 
of approximate testing equivalence under the three considered dimensions.
Moreover, as will be shown, in this framework it is possible to join the advantages of a decidable theory with the convenience 
of obtaining measures that can be easily interpreted in an activity oriented setting.

The remainder is organized as follows. First, we introduce some background about the testing framework (Sect.~\ref{sect:mtf}), 
i.e.\ we recall the Markovian process calculus and Markovian testing equivalence based on which we then formalize a notion of 
approximate testing equivalence from three different viewpoints (Sect.~\ref{sect:approx}).
The relaxed versions of Markovian testing equivalence based on time, probability, and observed behavior are presented separately 
and then combined in a unifying definition. Finally, the paper proposes some comparison with related work and interesting sights 
for future work (Sect.~\ref{sect:conc}).

%
\section{Markovian Testing Framework}\label{sect:mtf}
%

In this section, we recall Markovian testing equivalence in the setting of a Markovian process calculus that 
generates all the finite CTMCs with a minimum number of operators. For a complete survey of the main results concerning
these topics, the interested reader is referred to~\cite{ABC}.

\subsection{Markovian Process Calculus}

In the Markovian process calculus that we consider (MPC for short) every action is exponentially timed and its duration 
is described by a rate $\lambda \in \realns_{> 0}$ defining the exponential distribution such that the average duration 
of the action is given by the inverse of its rate. Formally, $\ms{Act} = \ms{Name} \times \realns_{> 0}$ is the set of 
actions of MPC, where $\ms{Name}$ is the set of action names, ranged over by $a,b,\ldots$, including the distinguished 
symbol $\tau$ denoting the invisible action.

The set of process terms of MPC is generated by the following syntax:
\[\begin{array}{l}
P ::= \nil \mid \lap a, \lambda \rap . P \mid P + P \mid A
\end{array}\]
where:
\begin{itemize}
\item The inactive process $\nil$ represents a terminated process.
\item The action prefix operator $\lap a, \lambda \rap . P$ represents a process performing the durational action 
$\lap a, \lambda \rap$ and then behaving as $P$.
\item The alternative composition operator encodes choice. If several durational actions can be performed the race policy 
is adopted, i.e.\ the fastest action is the one that is executed. The execution probability of each durational action is 
proportional to its rate and the average sojourn time associated with the related process term is exponentially distributed 
with rate given by the sum of the rates of the actions enabled by the term.
\item $A$ is a process constant defined by the possibly recursive equation $A \eqdef P$.
\end{itemize}

We denote with $\calp$ the set of closed and guarded process terms of MPC. The behavior of $P \in \calp$ is given by the labeled
multitransition system $\lsp P \rsp$, where states correspond to process terms and transitions are labeled with actions.
In particular, each transition has a multiplicity in order to keep track of the number of different proofs for the derivation of
the transition. This is necessary because the idempotent law does not hold in the stochastic setting. 
Indeed, a term like $\lap a, \lambda \rap . P + \lap a, \lambda \rap . P$ is not the same as $\lap a, \lambda \rap . P$ because 
of the race policy.

From the labeled multitransition system $\lsp P \rsp$ a CTMC can be easily derived by discarding the action names from the labels
and collapsing all the transitions between any pair of states into a single transition whose rate is the sum of the rates of the
collapsed transitions.

Formally, the semantic rules for MPC are as follows:
\[\begin{array}{c}
\langle a, \lambda \rangle . P \arrow{a,\lambda}{} P \\[2mm]
\infr{P_{1} \arrow{a, \lambda}{} P'}{P_{1} + P_{2} \arrow{a, \lambda}{} P'} \hspace{1cm}
\infr{P_{2} \arrow{a, \lambda}{} P'}{P_{1} + P_{2} \arrow{a, \lambda}{} P'} \\[8mm]
\infr{A \eqdef P \hspace{5mm} P \arrow{a,\lambda}{} P'}{A \arrow{a,\lambda}{} P'}
\end{array}\]

\subsection{Markovian Testing Equivalence}

Markovian testing equivalence is based on notions for process terms of MPC like exit rate -- the rate at which we leave the state
associated with the term -- and computation -- a sequence of transitions that can be executed starting from the state 
associated with the term. Below, we recall these two notions before introducing the testing scenario.

\begin{definition}
Let $P \in \calp$, $a \in \ms{Name}$, and $C \subseteq \calp$. 
The exit rate at which $P$ executes actions of name $a$ that lead to $C$ is defined through the non-negative real function:
\[
\begin{array}{c}
\ms{rate}(P, a, C) \: = \: 
\sum \lmp \lambda \in \realns_{> 0} \mid \exists P' \in C \ldotp P \arrow{a, \lambda}{} P' \rmp 
\end{array}\]
where the summation is taken to be zero whenever its multiset is empty.
\fullbox
\end{definition}

If we sum up the rates of all the actions that a process term $P$ can execute, we obtain the total exit rate of $P$.

	\begin{definition}

Let $P \in \calp$. The total exit rate of $P$ is defined as
$\ms{rate}_{\rm t}(P) \: = \: \sum\limits_{a \in \ms{Name}} \ms{rate}(P, a, \calp).$
\fullbox
	\end{definition}

The length of a computation is the number of transitions occurring in it. We denote with $\calc_{\rm f}(P)$ the multiset of 
finite-length computations of $P \in \calp$. Two distinct computations are independent of each other if neither is a proper 
prefix of the other one. In the remainder, we concentrate on finite sets of independent, finite-length computations.
We now define the concrete trace, the probability, and the duration of an element of $\calc_{\rm f}(P)$, using $\_ \circ \_$ 
for sequence concatenation and $| \_ |$ for sequence length.

	\begin{definition}

Let $P \in \calp$ and $c \in \calc_{\rm f}(P)$. The concrete trace associated with $c$ is the sequence of action names labeling 
the transitions of $c$, which is defined by induction on the length of $c$ through the $\ms{Name}^{*}$-valued function:
\[\begin{array}{c}
\ms{trace}(c) \: = \: \left\{ \begin{array}{ll}
\delta &
\hspace{0.5cm} \textrm{if $|c| = 0$} \\
a \circ \ms{trace}(c') &
\hspace{0.5cm} \textrm{if $c \equiv P \arrow{a, \lambda}{} c'$} \\
\end{array} \right. 
\end{array}\]
where $\delta$ is the empty trace.
\fullbox
	\end{definition}

	\begin{definition}

Let $P \in \calp$ and $c \in \calc_{\rm f}(P)$. The probability of executing $c$ is the product of the execution probabilities 
of the transitions of $c$, which is defined by induction on the length of $c$ through the $\realns_{]0, 1]}$-valued function:
\[\begin{array}{c}
\ms{prob}(c) \: = \: \left\{ \begin{array}{ll}
1 &
\hspace{0.5cm} \textrm{if $|c| = 0$} \\
{\lambda \over \ms{rate}_{\rm t}(P)} \cdot \ms{prob}(c') &
\hspace{0.5cm} \textrm{if $c \equiv P \arrow{a, \lambda}{} c'.$} \\
\end{array} \right. \\
\end{array}\]
The probability of executing a computation in $C \subseteq \calc_{\rm f}(P)$ -- whenever $C$ is finite and all of its 
computations are independent of each other -- is defined as:
\[\begin{array}{c}
\ms{prob}(C) \: = \: \sum\limits_{c \in C} \ms{prob}(c). 
\end{array}\]
\\[-14mm]
\fullbox
	\end{definition}

	\begin{definition}

Let $P \in \calp$ and $c \in \calc_{\rm f}(P)$. The stepwise average duration of $c$ is the sequence of average sojourn times 
in the states traversed by $c$, which is defined by induction on the length of $c$ through the $(\realns_{> 0})^{*}$-valued 
function:
\[\begin{array}{c}
\ms{time}(c) \: = \: \left\{ \begin{array}{ll}
\delta &
\hspace{0.5cm} \textrm{if $|c| = 0$} \\
{1 \over \ms{rate}_{\rm t}(P)} \circ \ms{time}(c') &
\hspace{0.5cm} \textrm{if $c \equiv P \arrow{a, \lambda}{} c'$} \\
\end{array} \right.
\end{array}\]
where $\delta$ is the empty stepwise average duration. We also define the multiset of computations in
$C \subseteq \calc_{\rm f}(P)$ whose stepwise average duration is not greater than $\theta \in (\realns_{> 0})^{*}$ as:
\[
\begin{array}{l}
C_{\le \theta} \: = \: \lmp c \in C \mid |c| \le |\theta| \land \forall i = 1, \dots, |c| \ldotp
\ms{time}(c)[i] \le \theta[i] \rmp.
\end{array}\]
Moreover, we denote by $C^{l}$ the multiset of computations in $C \subseteq \calc_{\rm f}(P)$ whose length
is equal to $l \in \natns$.
\fullbox
	\end{definition}

The main idea underlying the testing approach is that two process terms are equivalent whenever an external observer 
interacting with them by means of tests cannot infer any distinguishing information from the functional and quantitative 
standpoints. Tests are represented as process terms that interact with the terms to be tested through a parallel composition 
operator enforcing synchronization on all visible action names. A test is passed with success whenever a specific point 
during execution is reached.
In the rest of the paper, we model tests as non-recursive, finite-state process terms.

Intuitively, at each state the process term proposes the execution of a durational action chosen according to the race
policy and then, if such an action is visible, the test decides either to react by enabling the interaction or to block it
(note that tests cannot block the execution of $\tau$ actions). 
The interaction can occur between actions with the same name only. If the test offers several actions with the same name as 
that of the action chosen by the term, then the selection of one such actions is probabilistic.

Formally, tests consist of nondurational actions each equipped with a weight $w \in \realns_{>0}$. 
The set of tests respecting a canonical form is necessary and sufficient to decide whether two process terms are Markovian 
testing equivalent. Each of these canonical tests allows for one computation leading to success, whose intermediate 
states can have alternative computations leading to failure in one step. 

	\begin{definition}

The set $\tests_{\rm R,c}$ of canonical reactive tests is generated by the syntax:
\[\begin{array}{c}
T \: ::= \: \textrm{s} \mid \lap a, *_{1} \rap . T + \sum\limits_{b \in \cale - \{ a \}} \hspace{-0.2cm}
\lap b, *_{1} \rap . \textrm{f}
\end{array}\]
where 
$a \in \cale$, $\cale \subseteq \ms{Name} - \{\tau \}$ finite, the summation is absent whenever $\cale = \{ a \}$,
and \textrm{s} (resp.\ \textrm{f}) is a zeroary operator standing for success (resp.\ failure).
\fullbox
	\end{definition}

The following semantic rules define the interaction between a process term and a test:
\[\begin{array}{c}
\infr{P \arrow{\tau, \lambda}{} P'}
{P \pco{} T \arrow{\tau, \lambda}{} P' \pco{} T}
\hspace{1cm}
\infr{P \arrow{a, \lambda}{} P' \hspace{5mm} T \arrow{a, *_{w}}{} T'}
{P \pco{} T \llarrow{a, \lambda \cdot {w \over \ms{weight}(T,a)}}{} P' \pco{} T'} 
\end{array}\]
where $\ms{weight}(T,a) = \sum \lmp w \mid \exists T' \ldotp T \arrow{a, *_{w}}{\rm T} T' \rmp$ is the weight
of $T$ with respect to $a$ and $\arrow{}{\rm T}$ denotes the transition relation for tests.

Given $P \in \calp$ and $T \in \tests_{\rm R,c}$, the interaction system of $P$ and $T$ is the process term 
$P \pco{} T$, where each state of $\lsp P \pco{} T \rsp$ is called a configuration. We say that a configuration is 
successful if its test part is s and that a test-driven computation is successful if it traverses a successful 
configuration. 
We denote with $\calsc(P, T)$ the multiset of successful computations of $P \pco{} T$. It is worth noting that for any 
sequence $\theta \in (\realns_{> 0})^{*}$ of average amounts of time the multiset 
$\calsc_{\le \theta}^{|\theta|}(P, T)$ is finite and all the computations of it have a finite length and are independent 
of each other.

Markovian testing equivalence requires to compare the probabilities of performing successful test-driven computations 
within a given sequence of average amounts of time.

	\begin{definition}\label{mte}

Let $P_{1}, P_{2} \in \calp$. We say that $P_{1}$ is Markovian testing equivalent to $P_{2}$, written 
$P_{1} \sbis{\rm MT} P_{2}$, iff for all reactive tests $T \in \tests_{\rm R,c}$ and sequences 
$\theta \in (\realns_{> 0})^{*}$ of average amounts of time:
\cws{11}{\ms{prob}(\calsc_{\le \theta}^{|\theta|}(P_{1}, T)) \: = \: \ms{prob}(\calsc_{\le
\theta}^{|\theta|}(P_{2}, T)).}
\fullbox
	\end{definition}

The following example justifies why the average duration of a computation has been defined in terms of the sequence of 
average sojourn times in the states traversed by the computation, rather than simply considering the sum of average durations. 

\begin{example}
Consider the two process terms:
\cws{0}{\begin{array}{l}
\lap g, \gamma \rap . \lap a, \lambda \rap . \lap b, \mu \rap . \nil +
\lap g, \gamma \rap . \lap a, \mu \rap . \lap d, \lambda \rap . \nil \\
\lap g, \gamma \rap . \lap a, \lambda \rap . \lap d, \mu \rap . \nil +
\lap g, \gamma \rap . \lap a, \mu \rap . \lap b, \lambda \rap . \nil 
\end{array}}
Under the assumption $\lambda \not= \mu$ and $b  \not= d$, both terms have a computation with concrete trace $g \circ a \circ b$, 
probability $1 \over 2$, average duration ${1 \over 2 \cdot \gamma} + {1 \over \lambda} + {1 \over \mu}$, but 
different average sojourn times. We can argue similarly for the computation with concrete trace $g \circ a \circ d$. 
Intuitively, an external observer distinguishes between them by observing the names of the actions that are performed and 
the instants at which they are performed. This is captured by $\sbis{\rm MT}$ as the two process terms are not Markovian 
testing equivalent.
\fullbox

\end{example}

%
\section{Approximate Markovian Testing Equivalence}\label{sect:approx}
%

In this section, we show three levels of approximation for $\sbis{\rm MT}$.
The goal is to estimate from different perspectives how much a process term $P_2$ is similar to a given process term $P_1$. 
Here we assume that $P_1$ represents the original model to be approximated through an alternative model $P_2$.
Since similarity cannot be transitive, as usual when relaxing equivalence relations we will also investigate what can be 
``transitively'' inferred about the distance between two process terms $P_1$ and $P_3$ whenever there exists a process term 
$P_2$ that is similar to both of them.

The three considered dimensions of the similarity problem are: time taken to pass a test (Sect.~\ref{subsect:time}), expressed 
as the sequence of average sojourn times in the states traversed by successful computations; probability with which tests are 
passed (Sect.~\ref{subsect:prob}); syntactical form of the passed test (Sect.~\ref{subsect:test}). 
For every dimension, we will provide a measure of the distance between process terms that do not satisfy $\sbis{\rm MT}$,
by stepwise refining the notion of similarity in terms of flexibility and usability. 
In each case, we will discuss the interpretation of the measure and the complexity of the algorithm measuring the distance 
between process terms. Finally, we will present a unifying notion of approximate Markovian testing equivalence -- resulting 
in Def.~\ref{beh-mte-prob-time} -- which joins all the ingredients mentioned before. Indeed, a unifying framework is useful 
to study the trade-off existing among the three orthogonal aspects and the related impact upon the inequalities of the process 
terms under comparison. 

\subsection{Approximating Time}\label{subsect:time}

The first dimension under consideration is time. In the setting of $\sbis{\rm MT}$, the time needed to pass a test with
success is described as the sequence of average sojourn times in the states traversed by successful computations. 
Approximation at this level consists in relaxing the condition concerning the average sojourn times. 
We will introduce such an approximation through several steps in an incremental way. First, we will show how a process 
term $P_1$ can be approximated by a process term $P_2$ that is either ``slightly slower'' or ``slightly faster'' 
than $P_1$. Then, we join both interpretations of similarity in order to obtain the most general definition of Markovian 
testing similarity with respect to time.

We start by introducing the idea of slow approximation. Whenever $P_{2}$ approximates successful computations of $P_{1}$ with 
respect to a test $T$ and temporal threshold $\theta \in (\realns_{> 0})^{*}$, stepwise average sojourn times slightly greater 
than those imposed by $\theta$ may be tolerated. In this case, we obtain a slow approximation, in the sense that $P_2$ simulates 
$P_1$ -- the same tests are passed with the same probabilities -- but the successful computations of $P_2$ can be slower than 
the corresponding ones of $P_{1}$.

As a first attempt in formalizing this intuition, we define the multiset of computations in 
$C \subseteq \calc_{\rm f}(P)$ whose stepwise average duration is not greater than $\theta \in (\realns_{> 0})^{*}$ plus 
$\epsilon \in \realns_{\ge 0}$, which acts as a tolerance threshold:
\[
\begin{array}{l}
C_{\le \theta+\epsilon} \: = \: \lmp c \in C \mid |c| \le |\theta| \land \forall i = 1, \dots, |c| \ldotp 
\ms{time}(c)[i] \le \theta[i] + \epsilon \rmp. 
\end{array}\]
Based on this definition, we have the following relaxation of $\sbis{\rm MT}$.

	\begin{definition}\label{def1:sMts}

Let $P_{1}, P_{2} \in \calp$ and $\epsilon \in \realns_{\ge 0}$. 
We say that $P_{2}$ is slow Markovian testing $\epsilon$-similar to $P_{1}$ 
iff for all reactive tests $T \in \tests_{\rm R,c}$ and sequences $\theta \in (\realns_{> 0})^{*}$ of average 
amounts of time:
\cws{11}{\ms{prob}(\calsc_{\le \theta}^{|\theta|}(P_{1}, T)) \: = \: 
\ms{prob}(\calsc_{\le \theta+\epsilon}^{|\theta|}(P_{2}, T)).}
\fullbox
	\end{definition}

\begin{example}
Consider the process terms $P_{1} \eqdef \lap g, \gamma \rap . \lap a, \gamma \rap . \nil$ and 
$P_{2} \eqdef \lap g, \gamma-\delta \rap . \lap a, \gamma-\delta \rap . \nil$. Then, $P_{2}$ approximates 
(is slow Markovian testing $\epsilon$-similar to) $P_{1}$, where $\epsilon = {1 \over(\gamma-\delta)} - {1 \over \gamma}$ 
expresses exactly the difference between the stepwise average amounts of time of the computations of $P_{1}$ and $P_{2}$. 
\fullbox
\end{example}

\noindent Note that $P_1 \sbis{\rm MT} P_2$ if and only if $P_2$ (resp.\ $P_{1}$) is slow Markovian testing $0$-similar 
to $P_{1}$ (resp.\ $P_{2}$). Moreover, we have the following transitivity result.

\begin{proposition}\label{time:transitivity}

Let $P_{1}, P_{2}, P_{3} \in \calp$ and $\epsilon_{1}, \epsilon_{2} \in \realns_{\ge 0}$. 
If $P_{2}$ is slow Markovian testing $\epsilon_{1}$-similar to $P_{1}$ and $P_{3}$ is slow Markovian testing 
$\epsilon_{2}$-similar to $P_{2}$, then $P_{3}$ is slow Markovian testing $(\epsilon_{1}\!+\!\epsilon_{2})$-similar to $P_{1}$.
\fullbox

\end{proposition}

\noindent In favor of this approximation of $\sbis{\rm MT}$, we observe that it can be decided through a trivial variant of the 
algorithm for $\sbis{\rm MT}$ -- which will be outlined later in this section -- and with the same time complexity, which is $O(n^5)$, 
where $n$ is the total number of states of $\lsp P_{1} \rsp$ and $\lsp P_{2} \rsp$~\cite{ABC}. 
However, an approximation such as this is too restrictive, as illustrated in the following example.

\begin{example}

Consider the process terms of the previous example. Then, $P_{2}$ is not slow Markovian testing $\epsilon'$-similar 
to $P_{1}$, with $\epsilon' > {1 \over(\gamma-\delta)} - {1 \over \gamma}$. In fact, take $\theta[1]$ 
such that $\theta[1] < {1 \over \gamma} \;\wedge\; {1 \over \gamma-\delta} < \theta[1] + \epsilon'$. 
With this temporal threshold, any computation of $P_{1}$ is discarded, while this is not the case for $P_{2}$.
\fullbox
\end{example}

In order to further relax $\sbis{\rm MT}$, we need to compare explicitly the sets of computations of $P_{1}$ and $P_{2}$. 
Formally, given $C, C' \subseteq \calc_{\rm f}(P)$, we now define the multiset of computations in $C$ whose stepwise average 
duration is not greater than $\theta \in (\realns_{> 0})^{*}$ or else is $\epsilon$-similar, with $\epsilon \in \realns_{\ge 0}$, 
to the stepwise average duration of any computation in $C'_{\le \theta}$. Therefore:
\[\begin{array}{l}
C_{\le \theta+\epsilon,C'} \: = \: C_{\le \theta} \cup \\
\hspace{4mm}
\lmp c \in C \mid c \not\in C_{\le \theta} \wedge 
\exists c' \in C'_{\le \theta} \ldotp |c| \le |c'| \land \forall i = 1, \dots, |c| \ldotp 
\ms{time}(c')[i] \le \ms{time}(c)[i] \le \ms{time}(c')[i]+\epsilon \rmp.
\end{array}\]
Based on this definition, we propose a new approximation of $\sbis{\rm MT}$.

	\begin{definition}\label{def2:sMts}

Let $P_{1}, P_{2} \in \calp$ and $\epsilon \in \realns_{\ge 0}$. 
We say that $P_{2}$ is slow Markovian testing $\epsilon$-similar to $P_{1}$ 
iff for all reactive tests $T \in \tests_{\rm R,c}$ and sequences $\theta \in (\realns_{> 0})^{*}$ of average 
amounts of time:
\cws{11}{\ms{prob}(\calsc_{\le \theta}^{|\theta|}(P_{1}, T)) \: = \: 
\ms{prob}(\calsc_{\le \theta+\epsilon, \calsc^{|\theta|}(P_{1}, T)}^{|\theta|}(P_{2}, T)).}
\fullbox
	\end{definition}

\noindent Intuitively, $\calsc_{\le \theta}^{|\theta|}(P_{1}, T)$ is compared with $\calsc_{\le \theta}^{|\theta|}(P_{2}, T)$
augmented with the successful $T$-driven computations of $P_{2}$ that are slower (up to $\varepsilon$) 
than corresponding computations in $\calsc_{\le \theta}^{|\theta|}(P_{1}, T)$.

\begin{example}

Consider two process terms $P_{1}$ and $P_{2}$ that are defined as follows, respectively:
\cws{0}{\begin{array}{l}
\lap g, \gamma \rap . \lap a, \lambda \rap . \lap b, \lambda \rap . \nil +
\lap g, \gamma \rap . \lap a, \lambda \rap . \lap d, \lambda \rap . \nil \\
\lap g, \gamma \rap . \lap a, \lambda \rap . \lap d, \lambda-\delta \rap . \nil +
\lap g, \gamma \rap . \lap a, \lambda-\delta \rap . \lap b, \lambda \rap . \nil 
\end{array}}
The computation $c_{1}$ with concrete trace $g \circ a \circ b$ of $P_{1}$ is slowly $\epsilon$-simulated by the 
corresponding computation $c_{2}$ of $P_{2}$, provided that $\epsilon \ge {1 \over \lambda-\delta} - {1 \over \lambda}$. 
Given any test $T \in \tests_{\rm R,c}$, for each $\theta \in (\realns_{> 0})^{*}$ we have that 
$c_1 \in \calsc_{\le \theta}^{|\theta|}(P_{1}, T)$ iff
$c_2 \in \calsc_{\le \theta+\epsilon, \calsc^{|\theta|}(P_{1}, T)}^{|\theta|}(P_{2}, T)$, because from the temporal 
standpoint $c_2$ is stepwise slower than $c_1$ and their difference is limited by $\epsilon$. We can argue similarly in the 
case of the two computations with concrete trace $g \circ a \circ d$. Hence, $P_{2}$ is slow Markovian testing $\epsilon$-similar 
to $P_{1}$.
\fullbox
\end{example}

\noindent Note that $P_1 \sbis{\rm MT} P_2$ if and only if $P_2$ (resp.\ $P_{1}$) is slow Markovian testing $0$-similar to 
$P_{1}$ (resp.\ $P_{2}$). Moreover, we have the following transitivity result.

\begin{proposition}\label{time:transitivity_bis}

Let $P_{1}, P_{2}, P_{3} \in \calp$ and $\epsilon_{1}, \epsilon_{2} \in \realns_{\ge 0}$. 
If $P_{2}$ is slow Markovian testing $\epsilon_{1}$-similar to $P_{1}$ and $P_{3}$ is slow Markovian testing 
$\epsilon_{2}$-similar to $P_{2}$, then $P_{3}$ is slow Markovian testing $\delta$-similar to $P_{1}$ for some 
$\delta \le \epsilon_{1}+\epsilon_{2}$.
\fullbox
\end{proposition}

Alternatively, by a symmetric argument we obtain a fast approximation whenever the successful computations of 
$P_{1}$ are approximated by successful computations of $P_{2}$ with stepwise average duration that can be 
slightly lower than that of corresponding successful computations of $P_{1}$.
Based on this intuition, we have the following approximation of $\sbis{\rm MT}$ still preserving the same 
results concerning Def.~\ref{def2:sMts}.

	\begin{definition}\label{def:fMts}

Let $P_{1}, P_{2} \in \calp$ and $\epsilon \in \realns_{\ge 0}$. 
We say that $P_{2}$ is fast Markovian testing $\epsilon$-similar to $P_{1}$ 
iff for all reactive tests $T \in \tests_{\rm R,c}$ and sequences $\theta \in (\realns_{> 0})^{*}$ of average 
amounts of time:
\cws{11}{\ms{prob}(\calsc_{\le \theta+\epsilon, \calsc^{|\theta|}(P_{2}, T)}^{|\theta|}(P_{1}, T)) \: = \: 
\ms{prob}(\calsc_{\le \theta}^{|\theta|}(P_{2}, T)).}
\fullbox
	\end{definition}

\begin{example}

Consider a variant of the previous example where the second process term is:
\cws{0}{\begin{array}{l}
\lap g, \gamma \rap . \lap a, \lambda \rap . \lap d, \lambda+\delta \rap . \nil +
\lap g, \gamma \rap . \lap a, \lambda+\delta \rap . \lap b, \lambda \rap . \nil
\end{array}}
In this case, it is easy to see that $P_{2}$ is fast Markovian testing $\epsilon$-similar to $P_{1}$, where
$\epsilon \ge {{1 \over \lambda} - {1 \over \lambda + \delta}}$.
\fullbox
\end{example}

The definitions of (slow and fast) Markovian testing similarity can be decided in polynomial time by exploiting a simple 
variant of the same algorithm for $\sbis{\rm MT}$, because essentially the main objective -- i.e.\ equating the execution 
probability of certain successful computations -- does not change. The unique relaxation concerns the average durations of these 
computations, i.e.\ the criterion according to which the successful computations to compare are chosen.
We now outline the most important steps of this proof by illustrating the differences with respect to the original algorithm for
$\sbis{\rm MT}$ of~\cite{ABC}. First, deciding $\sbis{\rm MT}$ is reduced to decide the Markovian version 
of ready equivalence, which can be reduced to decide probabilistic ready equivalence if we consider the embedded discrete-time
versions of the CTMCs underlying the two process terms to compare. Then, probabilistic ready equivalence is decided through a 
suitable reworking of the algorithm for probabilistic language equivalence~\cite{Tzeng}. 
In the transformation from continuous time to discrete time, information about the total exit rate of each state is encoded 
within the action names labeling the transitions leaving that state. Note that the use of this additional information provides 
the unique difference between $\sbis{\rm MT}$ and (slow and fast) Markovian testing similarity. 
More precisely, when applying the algorithm for probabilistic language equivalence in the case of $\sbis{\rm MT}$, a state of 
$\lsp P_{1} \rsp$ is equated to a state of $\lsp P_{2} \rsp$, i.e.\ they are put into the same accepting set, if and only if 
the two sets of augmented action names labeling the transitions departing from the two states coincide. In particular, they must 
exhibit the same total exit rates. 
Hence, the temporal information represents a decoration that is used to decide which states of $\lsp P_{1} \rsp$ and 
$\lsp P_{2} \rsp$ belong to the same accepting set. In our relaxed setting, instead of checking the equality between the total 
exit rates as required by $\sbis{\rm MT}$, we check their inequality up to $\epsilon$, i.e.\ a state of $\lsp P_{1} \rsp$ is
equated to a state of $\lsp P_{2} \rsp$ if the total exit rate of the second state is greater/lower than the total exit rate of 
the first state and their difference is limited by the threshold $\epsilon$. 
Then, once the accepting sets are defined according to this condition, the algorithm of~\cite{Tzeng} proceeds as usual. 
The time complexity of the overall algorithm is $O(n^5)$.

Markovian testing similarity can be further relaxed. On the one hand, the fast and slow versions can be combined together, thus
obtaining the following definition.

	\begin{definition}\label{def:sfMts}

Let $P_{1}, P_{2} \in \calp$ and $\epsilon \in \realns_{\ge 0}$. 
We say that $P_{2}$ is temporally Markovian testing $\epsilon$-similar to $P_{1}$ 
iff for all reactive tests $T \in \tests_{\rm R,c}$ and sequences $\theta \in (\realns_{> 0})^{*}$ of average 
amounts of time:
\cws{11}{\ms{prob}(\calsc_{\le \theta+\epsilon, \calsc^{|\theta|}(P_{2}, T)}^{|\theta|}(P_{1}, T)) \: = \: 
\ms{prob}(\calsc_{\le \theta+\epsilon, \calsc^{|\theta|}(P_{1}, T)}^{|\theta|}(P_{2}, T)).}
\fullbox
	\end{definition}

\noindent Hence, a computation $c$ of $P_{1}$ can be approximated either by a slower or by a 
faster computation of $P_{2}$. However, $c$ cannot be approximated by a computation of $P_{2}$ that is stepwise 
either slower or faster than $c$. 
In order to overcome this limitation, we introduce the following relaxation of $C_{\le \theta+\epsilon,C'}$:
\[\begin{array}{l}
C_{\le \theta \pm \epsilon,C'} \: = \: C_{\le \theta} \cup \\
\hspace{1mm}
\lmp c \in C \mid c \not\in C_{\le \theta} \wedge 
\exists c' \in C'_{\le \theta} \ldotp |c| \le |c'| \land \forall i = 1, \dots, |c| \ldotp 
\ms{time}(c')[i]-\epsilon \le \ms{time}(c)[i] \le \ms{time}(c')[i]+\epsilon \rmp.
\end{array}\]
\noindent Based on this notion of approximation, a computation $c$ is similar to a computation $c'$ if the difference
between their average sojourn times is limited by $\epsilon$. Then, we have the following variant of Def.~\ref{def:sfMts}.

	\begin{definition}\label{def2:sfMts}

Let $P_{1}, P_{2} \in \calp$ and $\epsilon \in \realns_{\ge 0}$. 
We say that $P_{2}$ is temporally Markovian testing $\epsilon$-similar to $P_{1}$ 
iff for all reactive tests $T \in \tests_{\rm R,c}$ and sequences $\theta \in (\realns_{> 0})^{*}$ of average 
amounts of time:
\cws{11}{\ms{prob}(\calsc_{\le \theta \pm \epsilon, \calsc^{|\theta|}(P_{2}, T)}^{|\theta|}(P_{1}, T)) \: = \: 
\ms{prob}(\calsc_{\le \theta \pm \epsilon, \calsc^{|\theta|}(P_{1}, T)}^{|\theta|}(P_{2}, T)).}
\fullbox
	\end{definition}

\noindent Note that this extension does not alter the decidability results of Markovian testing similarity.

\begin{example}

Consider a variant of the previous example where the second process term is:
\cws{0}{\begin{array}{l}
\lap g, \gamma \rap . \lap a, \lambda-\delta \rap . \lap d, \lambda+\delta \rap . \nil +
\lap g, \gamma \rap . \lap a, \lambda+\delta \rap . \lap b, \lambda-\delta \rap . \nil
\end{array}}
It can be verified that $P_{2}$ is temporally Markovian testing $\epsilon$-similar to $P_{1}$, where
$\epsilon \ge {{1 \over \lambda - \delta} - {1 \over \lambda}}$.
\fullbox
\end{example}

On the other hand, when comparing the computations of two process terms we can decide to change at each step the value 
of the threshold expressing the tolerance to different temporal behaviors. This is obtained by assuming 
$\epsilon \in (\realns_{\ge 0})^{*}$ and checking, e.g., the inequality:
\[
\begin{array}{l}
\forall i = 1, \dots, |c| \ldotp \ms{time}(c')[i] \le \ms{time}(c)[i] \le \ms{time}(c')[i]+\epsilon[i] 
\end{array}\]
within the definition of $C_{\le \theta+\epsilon,C'}$. For instance, this variant can be used to discount the effect 
of far (in the future) steps by assuming that $\epsilon[i]$ increases as long as $i$ increases.

\subsection{Approximating Probability}\label{subsect:prob}

The introduction of a relaxation concerning the probabilistic behavior of process terms results into the following extension of 
$\sbis{\rm MT}$ where the probabilities of the successful $T$-driven computations of $P_{1}$ and $P_{2}$ are not imposed to be 
equal anymore.

	\begin{definition}\label{def:pMts}

Let $P_{1}, P_{2} \in \calp$ and $\epsilon \in \realns_{\ge 0}$. 
We say that $P_{2}$ is probabilistically Markovian testing $\epsilon$-similar to $P_{1}$ 
iff for all reactive tests $T \in \tests_{\rm R,c}$ and sequences $\theta \in (\realns_{> 0})^{*}$ of average 
amounts of time:
\cws{11}{|\ms{prob}(\calsc_{\le \theta}^{|\theta|}(P_{1}, T)) - \ms{prob}(\calsc_{\le \theta}^{|\theta|}(P_{2}, T))|
\le \epsilon.}
\fullbox
	\end{definition}

As we have seen in the previous section, verifying Markovian testing equivalence amounts to decide whether two probabilistic 
automata accept the same words with the same probability. However, as shown in~\cite{dRT}, the relaxation of this equivalence
problem, i.e.\ checking whether for all words the distance between two process models is less than $\epsilon$, is an
undecidable problem.

To make it decidable, it is possible to restrict ourselves to more specific notions of probabilistic similarity. 
As an example, \cite{ST} defines a polynomially accurate similarity that can be rephrased in our testing framework as follows: any 
set of successful computations of $P_1$ with a polynomial number of steps must be matched by $P_2$ with an error that is bounded 
by any polynomial. In order to measure the distance between process terms even when their difference is not negligible in the sense
of~\cite{ST}, in the next section we will show that decidability is obtained by relaxing the condition over tests in Def.~\ref{def:pMts}.

\subsection{Approximating Tests}\label{subsect:test}

Similarly as done in Sect.~\ref{subsect:time}, in this section we consider in an incremental way a notion of similarity that is based 
on the exemplary behavior of tests. The proposed approach is not completely naive as it is somehow inspired by~\cite{AAW}, where processes 
are compared with respect to an event log describing typical behaviors. In particular, in~\cite{AAW} processes are defined in terms of 
Petri nets and an event log is a multiset of firing sequences. Then, different models are compared by measuring the overlap in (partially) 
fitting these sequences. 
This is done by using a fitness function that takes into account all enabled transitions at any point in each sequence. 
This idea results into two measures, called precision and recall. Precision establishes whether the behavior of the second, 
alternative model is possible from the viewpoint of the behavior of the first, original model. Recall establishes how much of 
the behavior of the first model is covered by the second model. 
In our setting, we resort to a variant of this kind of approach from two different perspectives. 

First, we observe that the notion of typical behavior that is at the base of model evaluation is naturally represented by tests.
While in~\cite{AAW} it is suggested to define the event log through simulation or by explicitly describing by hand some 
typical behavior of interest, in our setting we formally describe an event log as a finite set of tests satisfying properties 
described in terms of logical formulas. Canonical tests do not exhibit any probabilistic and temporal behavior, so that  
we can employ the logical characterization of testing equivalence, which comprises a restricted set of logical 
operators: a modal operator on sequences of visible actions, true, disjunction, and diamond~\cite{ABC}. Then, given a formula 
$\phi$ representing a property of interest, we use as event log the set of canonical tests satisfying $\phi$, called 
$\tests_{\rm R,c,\phi}$, provided that such a set is finite. As an example, $\phi$ could be the formula 
that is satisfied by all the tests in which the unique computation leading to success is made of the concrete trace 
$a_1 \circ \ldots \circ a_n$. Thus, this trace represents the property with respect to which it is interesting to compare two 
process terms.
In general, tests satisfying $\phi$ denote the set of typical behaviors, parameterized by $\phi$, which guide the estimation of 
the degree of similarity between process terms.

Second, we observe that a test-based notion of the fitness measures of~\cite{AAW} can be used to estimate the similarity between
tests. Approximating tests, as well as relaxing time and probability requirements, is justified by the fact that we intend to
overcome the typical limitations of ``perfect'' equivalence.
In order to relax $\sbis{\rm MT}$ by following this intuition, we assume that the process terms to compare are not expected to 
exhibit the same quantitative behavior when interacting with the same test, but they can exhibit such a behavior when 
interacting with two possibly different but similar tests. 
In other words, if a process term satisfies a test with a certain probability and within a given amount of time, then the second 
one can simulate the behavior of the first term by satisfying with the same probability and by the same time another test that 
fits the first test according to a notion of test similarity.

Inspired by the formulas of~\cite{AAW}, we now define the notions of behavioral precision and recall for test similarity. 
Let $\ms{trace}_{s}(T)$ be the concrete trace associated with the unique computation of $T$ leading to success, $|T|$ be 
the length of this trace, and $T_{i}$ be the i-th state of it, such that $T_{1} ::= T$ and $T_{|T|}$ is the state 
that reaches success in one step. Then, we assume that $\forall i = 1, \ldots, |T|$, $\ms{enabled}(T,i,s) = a$ iff 
$\ms{trace}_{s}(T)[i] = a$ and
$\ms{enabled}(T,i,f) = \{ b\,\mid\, T_{i} ::= \lap b, *_{1} \rap . \mathrm{f} + T' \}$. 
In practice, $\ms{enabled}(T,i,s)$ denotes the transition belonging to the successful computation of $T$ that is enabled at the 
$i$-th step, while $\ms{enabled}(T,i,f)$ denotes the set of transitions leading to failure in one step that are enabled at the 
$i$-th step. Then, we introduce the following definitions of precision and recall for two tests $T$ and $T'$:\\
\cws{0}{
\begin{array}{rcl}
\ms{prec}(T,T') & = & {1 \over \mid T' \mid} \sum_{i=1}^{\mid T' \mid} 
{ \mid (\ms{enabled}(T,i,s) \,\cap\, \ms{enabled}(T',i,s)) \;\cup\; (\ms{enabled}(T,i,f) \,\cap\, \ms{enabled}(T',i,f)) \mid
\over 
\mid \ms{enabled}(T',i,f) \mid \,+\, \mid \ms{enabled}(T',i,s) \mid}
\end{array}}
and: \\
\cws{0}{
\begin{array}{rcl}
\ms{rec}(T,T') & = & {1 \over \mid T \mid} \sum_{i=1}^{\mid T \mid} 
{ \mid (\ms{enabled}(T,i,s) \,\cap\, \ms{enabled}(T',i,s)) \;\cup\; (\ms{enabled}(T,i,f) \,\cap\, \ms{enabled}(T',i,f)) \mid
\over 
\mid \ms{enabled}(T,i,f) \mid \,+\, \mid \ms{enabled}(T,i,s) \mid}.\\
\end{array}}

At each step, we compare the set of enabled transitions for the current state of the two tests, by distinguishing 
the transitions leading to failure from the unique one along the computation leading to success. 
Both formulas establish a measure between $0$ and $1$ that estimates the similarity between them. Obviously, it holds that 
$\ms{prec}(T,T') = \ms{rec}(T',T)$.
Similarly as in~\cite{AAW}, it is important to note that tests are not imposed to offer the same behavior, which may differ 
step by step thus originating different computations.

Analogously, $T$ and $T'$ are not imposed to have the same length. For instance, if $|T|= 2 \cdot |T'| = 2 \cdot n$ and 
the behaviors of $T$ and $T'$ coincide in the first $n$ steps, then $\ms{prec}(T,T') = 1$ because each behavior of $T'$ is 
possible according to the behavior of $T$, while $\ms{rec}(T,T') = {1 \over 2}$ because only half of the behavior of $T$ is 
covered by the behavior of $T'$. 
On the other hand, $T$ and $T'$ coincide iff $\ms{prec}(T,T') = \ms{rec}(T,T') = 1$.

\begin{example}

Consider $T_1 = \lap a, *_{1} \rap . \mathrm{s} + \lap b, *_{1} \rap . \mathrm{f}$ and 
$T_2 = \lap b, *_{1} \rap . \mathrm{s} + \lap a, *_{1} \rap . \mathrm{f}$. Then, it holds that 
$\ms{prec}(T_{1},T_{2}) = \ms{rec}(T_{1},T_{2}) = 0$ because we distinguish actions leading to success from those leading 
to failure. Without this distinction, it would result $\ms{prec}(T_{1},T_{2}) = \ms{rec}(T_{1},T_{2}) = 1$.

Now, consider the two tests 
$T_{1} = \lap a_{1}, *_{1} \rap . \lap a_{2}, *_{1} \rap . \mathrm{s} + \lap b, *_{1} \rap . \mathrm{f}$ and 
$T_{2} = \lap c, *_{1} \rap . \lap a_{2}, *_{1} \rap . \mathrm{s} + \lap b, *_{1} \rap . \mathrm{f} + 
\lap b', *_{1} \rap . \mathrm{f}$. Then, $\ms{prec}(T_{1},T_{2}) = {2 \over 3}$ and $\ms{rec}(T_{1},T_{2}) = {3 \over 4}$.
Recall is higher than precision, because the unique behavior of $T_{1}$ that is not covered by $T_{2}$ is the first action 
of the successful computation, while from the viewpoint of $T_{1}$ we have two impossible behaviors of $T_{2}$, i.e.\
the actions $c$ and $b'$.
\fullbox
\end{example}

Precision and recall satisfy the same transitivity relations shown in~\cite{AAW}, as reported in 
Table~\ref{prec_rec_transitive} for the sake of completeness.

\begin{table}
\footnotesize
\[\begin{array}{|c|c|c|c||c|c|}
\hline 
\ms{prec(T_{1},T_{2})} & \ms{rec(T_{1},T_{2})} & \ms{prec(T_{2},T_{3})} & \ms{rec(T_{2},T_{3})} & \ms{prec(T_{1},T_{3})} 
& \ms{rec(T_{1},T_{3})} \\
z & w & x & y & \le 1 & \le 1 \\
z & w & x & 1 & < 1   & \ge w \\
z & w & 1 & y & \le 1 & \le w \\
z & w & 1 & 1 & z     & w     \\
z & 1 & x & y & \le x & \le 1 \\
z & 1 & x & 1 & < x   & 1     \\
z & 1 & 1 & y & \le 1 & \le 1 \\
z & 1 & 1 & 1 & z     & 1     \\
1 & w & x & y & \ge x & \le 1 \\
1 & w & x & 1 & \ge x & \ge w \\
1 & w & 1 & y & 1     & < w   \\
1 & w & 1 & 1 & 1     & w     \\
1 & 1 & x & y & x     & y     \\
1 & 1 & x & 1 & x     & 1     \\
1 & 1 & 1 & y & 1     & y     \\
1 & 1 & 1 & 1 & 1     & 1     \\
\hline
\end{array}\]
\caption{Transitivity relations for $\ms{prec}$ and $\ms{rec}$: $z,w,x,y \in [0,1[$}\label{prec_rec_transitive}
\end{table}

Then, by using a notion of test similarity quantified with respect to the precision and recall defined above, we have the 
following relaxation of $\sbis{\rm MT}$, which is based on the observed behavior expressed in terms of test-driven 
computations, where instead of a single test we consider a pair of tests that fit almost the same. 
The first attempt abstracts from the temporal behavior of the process terms to compare.

	\begin{definition}\label{def1:bMts}

Let $P_{1}, P_{2} \in \calp$ and $\tests_{\rm R,c,\phi}$ a finite set of tests. 
We say that $P_{2}$ is behaviorally Markovian testing similar to $P_{1}$ with precision $p \in [0,1]$ and recall 
$r \in [0,1]$ iff for each reactive test $T \in \tests_{\rm R,c,\phi}$ there exists a reactive test 
$T' \in \tests_{\rm R,c,\phi}$ such that:
\begin{enumerate}
\item $\ms{prec(T,T')} \ge p$ and $\ms{rec(T,T')} \ge r$
\item $\ms{prob}(\calsc(P_{1}, T)) \: = \: \ms{prob}(\calsc(P_{2}, T')).$
\fullbox
\end{enumerate}

	\end{definition}

As far as the transitivity properties of Def.~\ref{def1:bMts} are concerned, we now discuss what can be inferred about two 
process terms $P_{1}$ and $P_{3}$ provided that there exists a process term $P_{2}$ such that $P_{2}$ is behaviorally 
Markovian testing similar to $P_{1}$ with precision $p$ and recall $r$ and $P_{3}$ is behaviorally Markovian testing similar 
to $P_{2}$ with precision $p'$ and recall $r'$.
By hypothesis, for each test $T$ applied to $P_{1}$ there exists a test $T'$ applied to $P_{2}$ such that the probabilities 
of the successful $T$-driven computations of $P_{1}$ and of the successful $T'$-driven computations of $P_{2}$ are equal. 
By hypothesis, there exists also a test $T''$ applied to $P_{3}$ such that the probabilities of the successful $T'$-driven  
computations of $P_{2}$ and of the successful $T''$-driven computations of $P_{3}$ are equal. Hence, the probabilities of the 
successful $T$-driven computations of $P_{1}$ and of the successful $T''$-driven computations of $P_{3}$ are equal. 
Afterwards, $\ms{prec}(T,T'')$ and $\ms{rec}(T,T'')$ can be inferred from $p$, $r$, $p'$, and $r'$, as shown in 
Table~\ref{prec_rec_transitive}.

In order to take into different account behaviors with a very low probability of success in comparison with successful 
behaviors occurring more frequent, in the two inequalities of Def.~\ref{def1:bMts} we can multiply $p$ and $r$ by the 
probability of the successful test-driven computations of $P_{1}$.

The next step refines the condition about probabilities of Def.~\ref{def1:bMts} by taking into account the temporal behavior of 
process terms. 
We recall that $\sbis{\rm MT}$ is defined with respect to all the sequences $\theta \in (\realns_{> 0})^{*}$ of average amounts 
of time. When considering a canonical test $T$ and a process term $P$ that does not execute invisible actions, we can restrict 
ourselves to the sequences of length $|T|$, which is the exact number of steps needed to reach success. This is not enough to 
reduce the comparison between $T$ and a similar test $T'$ to a finite set of sequences. 
Therefore, we now define a canonical set of sequences for $T$ that is finite and is sufficient to decide whether 
a process term behaviorally simulates another one with respect to $T$.

Such a canonical set is made of a sequence for each subset of the set of successful computations $\calsc^{|T|}(P,T)$. 
For each $X \in 2^{\calsc^{|T|}(P,T)}$ we define the sequence of average amounts of time $\theta_{X}$ such that 
$\forall i = 1,\ldots,|T| \ldotp \theta_{X}[i] = \max_{c \in X}\{\ms{time}(c)[i]\}$ and the canonical set 
$\Theta(P,T) = \{ \theta_{X} \,\mid\, X \in 2^{\calsc^{|T|}(P,T)} \}$.
Note that $X \subseteq \calsc_{\le \theta_{X}}^{|T|}(P,T)$ and that we may have 
$\calsc_{\le \theta_{X}}^{|T|}(P,T) = \calsc_{\le \theta_{Y}}^{|T|}(P,T)$ for some $X \not= Y$, so that the minimum number
of sequences to consider could be lower than $|2^{\calsc^{|T|}(P,T)}|$.

The algorithm that computes these sequences consists of building a tree as follows. The root is at level $1$ and is marked 
with the set of all the successful computations $\calsc^{|T|}(P,T)$. If the current node of the level $i$ is marked with a 
set $\cals$ of computations, then create a child node for each $Y \subseteq \cals$ for which there exists $k \in \realns_{> 0}$ 
such that $\ms{times}(c)[i] \le k$ for each $c \in Y$. 
Add to this new node the labels $Y$ and $\max_{c \in Y}\{\ms{time(c)[i]}\}$. The tree
construction terminates at the level $|T|+1$. In this way, the tree contains at most $|2^{\calsc^{|T|}(P,T)}|$ leafs,
each leaf is associated with a subset $X \in 2^{\calsc^{|T|}(P,T)}$, and the path from the root to this leaf contains as labels
the average amounts of time forming the sequence $\theta_{X}$.

\begin{proposition}\label{canonical_sequences}

Let $P_{1}, P_{2} \in \calp$ and $T \in \tests_{\rm R,c}$. 
If for each sequence $\theta \in \Theta(P_{1},T) \cup \Theta(P_{2},T)$ of average amounts of time we have: 
\cws{0}{\ms{prob}(\calsc_{\le \theta}^{|\theta|}(P_{1}, T)) \: = \: 
\ms{prob}(\calsc_{\le \theta}^{|\theta|}(P_{2}, T))}
then, we also have that for each sequence $\theta \in (\realns_{> 0})^{*}$ of average 
amounts of time: 
\cws{11}{\ms{prob}(\calsc_{\le \theta}^{|\theta|}(P_{1}, T)) \: = \: 
\ms{prob}(\calsc_{\le \theta}^{|\theta|}(P_{2}, T)).}
\fullbox

\end{proposition}

\noindent Now, we are ready to define a decidable approximation of $\sbis{\rm MT}$ based on observed behavior.

	\begin{definition}\label{def2:bMts}

Let $P_{1}, P_{2} \in \calp$ and $\tests_{\rm R,c,\phi}$ a finite set of tests. 
We say that $P_{2}$ is behaviorally Markovian testing similar to $P_{1}$ with precision $p \in [0,1]$ and recall 
$r \in [0,1]$ iff for each reactive test $T \in \tests_{\rm R,c,\phi}$ there exists a reactive test 
$T' \in \tests_{\rm R,c,\phi}$ such that for all sequences $\theta \in \Theta(P_{1},T) \cup \Theta(P_{2},T')$ 
of average amounts of time:
\begin{enumerate}
\item $\ms{prec(T,T')} \ge p$ and $\ms{rec(T,T')} \ge r$
\item $ \ms{prob}(\calsc_{\le \theta}^{|\theta|}(P_{1}, T)) \: = \: \ms{prob}(\calsc_{\le \theta}^{|\theta|}(P_{2}, T')).$
\fullbox
\end{enumerate}

	\end{definition}

The same considerations concerning the transitivity of Def.~\ref{def1:bMts} still hold.
With respect to the approximations based on time and probability that have been discussed in the previous sections,
in this setting we deal with finite sets of tests and sequences of average amounts of time. Hence, it is possible to define 
a very intuitive, still decidable, approximation of $\sbis{\rm MT}$ based on time, probability, observed behavior, and the 
three corresponding families of quantitative thresholds.

	\begin{definition}\label{beh-mte-prob-time}

Let $P_{1}, P_{2} \in \calp$ and $\tests_{\rm R,c,\phi}$ a finite set of tests. 
We say that $P_{2}$ is Markovian testing similar to $P_{1}$ with precision $p \in [0,1]$, recall 
$r \in [0,1]$, temporal threshold $\epsilon \in \realns_{> 0}$, and probability threshold $\nu \in \realns_{> 0}$ 
iff for each reactive test $T \in \tests_{\rm R,c,\phi}$ there exists a reactive test 
$T' \in \tests_{\rm R,c,\phi}$ such that for all sequences $\theta \in \Theta(P_{1},T) \cup \Theta(P_{2},T')$ 
of average amounts of time:
\begin{enumerate}
\item $\ms{prec(T,T')} \ge p$ and $\ms{rec(T,T')} \ge r$
\item $ |\ms{prob}(\calsc_{\le \theta \pm \epsilon, \calsc^{|\theta|}(P_{2}, T')}^{|\theta|}(P_{1}, T)) - 
\ms{prob}(\calsc_{\le \theta \pm \epsilon, \calsc^{|\theta|}(P_{1}, T)}^{|\theta|}(P_{2}, T'))| \le \nu.$
\fullbox
\end{enumerate}

	\end{definition}

\noindent Given a modal logic formula $\phi$, we observe that $P_2$ (resp.\ $P_{1}$) is Markovian testing similar to $P_{1}$ 
(resp.\ $P_{2}$) with precision 1, recall 1, temporal and probability thresholds 0 if and only if $P_1 \sbis{\rm MT} P_2$ 
with respect to the tests defined by $\phi$.  It is worth noting that a unifying framework merging the three orthogonal 
aspects (time, probability, and observed behavior) puts the basis for the analysis of the trade-off among them.

\begin{example}
Consider two process terms $P_{1}$ and $P_{2}$ that are defined as follows, respectively:
\cws{0}{\begin{array}{l}
\lap g, \gamma \rap . \lap a, \lambda+\delta \rap . \lap b, \lambda \rap . \nil +
\lap g, \gamma \rap . \lap a, \lambda \rap . \lap d, \lambda \rap . \nil \\
\lap g, \gamma \rap . \lap a, \lambda \rap . \lap d', \lambda \rap . \nil +
\lap g, \gamma \rap . \lap a, \lambda \rap . \lap b, \lambda-\delta \rap . \nil 
\end{array}}
and compare them with respect to tests whose successful computation is described by the concrete trace $g \circ a \circ \ast$,
with $\ast$ any action. Then, $P_{2}$ is Markovian testing similar to $P_{1}$ with:
\begin{itemize}
\item both precision and recall equal to $2 \over 3$, where the difference in the observed behaviors is due to the two concrete 
traces $g \circ a \circ d$ of $P_{1}$ and $g \circ a \circ d'$ of $P_{2}$, under the assumption $d \not= d'$; 
\item temporal threshold $\epsilon \ge {1 \over \lambda - \delta} - {1 \over \lambda} > 
{1 \over \lambda} - {1 \over \lambda + \delta}$, where the difference in the average 
sojourn times is due to the three rates $\lambda$, $\lambda+\delta$, $\lambda-\delta$ labeling corresponding transitions
related to the two concrete traces $g \circ a \circ b$ of $P_{1}$ and $P_{2}$; 
\item probability threshold $0$, since the probabilities of the successful computations to compare are always the same.
\fullbox
\end{itemize}
\end{example}

%
\section{Related and Future Work}\label{sect:conc}
%

In the last decade several approaches to the approximation of behavioral equivalences have been proposed (see, e.g., 
\cite{LMMS,DGJP,vBW,DHW,AP,AAW,DLT,DHW2} and the references therein). 

Some of them use a well-established approach based on behavioral pseudometrics~\cite{DGJP,vBW}, which give a 
measure of the similarity between states of a transition system. These pseudometrics provide a conservative extension of 
bisimulation equivalence. Hence, they cannot be compared with the notions of testing similarity, which instead rely on testing 
semantics. With approaches based on pseudometrics it is not easy to establish a clear relation between the measure estimating 
process similarity and its interpretation in a practical, mainly activity oriented, setting. 
As an example elucidating this aspect, \cite{AB} shows the importance of evaluating the impact that the absence of an equivalence 
relating two process terms has upon their difference with respect to performability measures. However, this is done without 
defining explicitly an approximate equivalence relating these measures with the degree of similarity. 

Some other approaches that are not based on pseudometrics, like~\cite{AP,DLT,DHW2}, rely on relations approximating 
bisimulation equivalence. 
These approaches seem promising thanks to the strict relation between bisimulation and lumping for Markov chains~\cite{Buc}. 
Indeed, the characterization of lumpability is extremely useful, because the knowledge of a lumpable partition of the states 
of a Markov chain allows the generation of an aggregated Markov chain that is smaller than the original one, but leads to 
several results for the original Markov chain without an error. In this setting, there exist approximation techniques based 
on relaxed notions of lumping and on perturbation theory which establish bounds on the error made when approximating. This is 
particularly useful because these bounds are in direct relation with the numerical analysis of Markov chains and, therefore, 
provide immediately a clear interpretation of their impact upon the quantitative behavior of the process terms under analysis. 
However, it seems that there still exists a significant gap between the applicability of the approximate bisimulations 
mentioned above and their decidability. Very often, the (strict) assumptions underlying approximate bisimulation that are needed 
to define efficient verification algorithms are such that it becomes hard to find real application domains and, in particular, 
to give a natural interpretation of the degree of similarity. 
On the other hand, the definition of an approximate bisimulation that can be related to approximate lumping and has an efficient 
verification algorithm is still an open problem.

Contrariwise, the approach proposed in~\cite{AAW} does not rely on behavioral equivalences, since it is based on the estimation
of observed behaviors -- quantified through a notion of fitness that does not require any nonfunctional information such as time 
and probability -- whenever log-driven computations are compared. However, this estimation is not related to any notion of
behavioral equivalence.

The main result of this paper is showing that testing equivalence offers an ideal semantic framework for joining ideas
taken from approximate behavioral equivalences with the approach of~\cite{AAW}. In addition, the proposed definitions of 
approximation elucidate the role of each aspect under consideration -- time, probability, and observed behavior -- without 
sacrificing neither decidability nor usability. 

As future work, it would be interesting to investigate the relation between the estimations provided by approximate 
Markovian testing equivalence and $T$-lumpability~\cite{ABC}, which is the version of lumpability corresponding to Markovian 
testing equivalence. One such result would enhance the applicability to domains where the degree of similarity must be 
interpreted in terms of impact upon the performance behavior of systems.

The application to real examples will be the subject of further investigations. For instance, it is well-known that approximate
equivalence checking can be profitably employed in the setting of noninterference analysis. Basically, one user/component 
may affect the behavior of other users/components in a way that compromises system properties like security and safety. 
Such an impact is studied by comparing the two views of the system that are obtained by activating and deactivating, respectively, 
the behavior of the interfering user/component. This approach is illustrated and used in~\cite{ABC} for the evaluation of 
performability aspects of several real-world case studies, like a secure routing system and a power-manageable system.
In this setting, the goal is to use Markovian testing similarity to compare different system views with respect to families of 
properties formalized through modal logic formulas. The comparison is intended to distinguish which observable 
behaviors make these views different from functional, temporal, and probabilistic perspectives, each case accompanied by a 
measure of such a difference.

\medskip
\noindent
 
        {\bf Acknowledgement}
 
\noindent
The author thanks the anonymous referees for their valuable comments. 
This work has been funded by MIUR-PRIN project \textit{PaCo -- Performability-Aware Computing: Logics, Models, and Languages}.

\bibliographystyle{eptcs} 

\end{document}

%% file: Environments.tex
\newtheorem{new_theorem}
	{Theorem}[section]

\newtheorem{new_definition}
	[new_theorem]{Definition}

\newtheorem{new_remark}
	[new_theorem]{Remark}

\newtheorem{new_example}
	[new_theorem]{Example}

\newtheorem{new_lemma}
	[new_theorem]{Lemma}

\newtheorem{new_proposition}
	[new_theorem]{Proposition}

\newtheorem{new_corollary}
	[new_theorem]{Corollary}

\newenvironment{definition}
	{\begin{new_definition}\rm}
	{\end{new_definition}}

\newenvironment{example}
	{\begin{new_example}\rm}
	{\end{new_example}}

\newenvironment{proposition}
	{\begin{new_proposition}\rm}
	{\end{new_proposition}}



%% file: Commands.tex
\def\ms#1{\null\ifmmode\mathord{\mathcode`-="702D\it #1\mathcode`\-="2200}%
	\else$\mathord{\mathcode`-="702D\it #1\mathcode`\-="2200}$\fi}

\newcommand{\cws}[2]
	{\\ \centerline{$#2$} \\[-#1pt]}

\newlength{\spacelen}

\newcommand{\lap}
	{\mbox{$<$}}
\newcommand{\rap}
	{\mbox{$>$}}

\newcommand{\lsp}
	{[ \! [}
\newcommand{\rsp}
	{] \! ]}

\newcommand{\lmp}
	{\{ \! | \,}
\newcommand{\rmp}
	{\, | \! \}}

\newcommand{\calc}
        {\mathcal{C}}

\newcommand{\cale}
        {\mathcal{E}}

\newcommand{\calp}
        {\mathcal{P}}

\newcommand{\cals}
        {\mathcal{S}}

\newcommand{\calsc}
        {\mathcal{SC}}

\newcommand{\natns}
	{\mathbb{N}}

\newcommand{\realns}
	{\mathbb{R}}

\newcommand{\tests}
	{\mathbb{T}}

\newcommand{\arrow}[2]
        {\, {\auxarrow\limits^{#1}}_{#2} \,}
\newcommand{\auxarrow}
	{\mathop{- \!\! - \!\!\!\! \longrightarrow}}

\newcommand{\llarrow}[2]
        {\, {\auxllarrow\limits^{#1}}_{#2} \,}
\newcommand{\auxllarrow}
        {\mathop{- \hspace{-0.2cm} - \hspace{-0.2cm} - \hspace{-0.2cm}
	- \hspace{-0.2cm} - \hspace{-0.2cm} - \hspace{-0.2cm}
	- \hspace{-0.2cm} - \hspace{-0.3cm} \longrightarrow}}

\newcommand{\nil}
	{\underline 0}

\newcommand{\eqdef}
	{\buildrel \Delta \over =}

\newcommand{\sbis}[1]
	{\sim_{#1}}

\newcommand{\pco}[1]
	{\mathop{\Vert_{#1}}}

\newcommand{\infr}[2]
	{\renewcommand{\arraystretch}{1.5}
	\begin{array}{c}
	#1\\
	\hline
	#2
	\end{array}}

\newcommand{\fullbox}
	{{\mbox{}\nolinebreak\hfill{$\rule{2mm}{2mm}$}}}





%% file: qapl10.bbl
\begin{thebibliography}{1}

\bibitem{AABBBL}
A.~Acquaviva, A.~Aldini, M.~Bernardo, A.~Bogliolo, E.~Bont\`a, and E.~Lattanzi (2005):
\newblock \emph{A Formal Method Based Methodology for Predicting the Impact of Dynamic Power Management}.
\newblock {\sl Formal Methods for Mobile Computing, Springer LNCS} 3465, pp.~155--189. 

\bibitem{ABC}
A.~Aldini, M.~Bernardo, and F.~Corradini (2010):
\newblock \emph{A Process Algebraic Approach to Software Architecture Design}.
\newblock {\sl Springer}.

\bibitem{AB}
A.~Aldini and M.~Bernardo (2009):
\newblock \emph{Weak Behavioral Equivalences for Verifying Secure and Performance-Aware Component-Based Systems}.
\newblock {\sl Architecting Dependable Systems~6, Springer LNCS} 5835, pp.~228--254. 

\bibitem{AP}
A.~Aldini and A.~Di~Pierro (2008):
\newblock \emph{Estimating the Maximum Information Leakage}.
\newblock {\sl Journal of Information Security} 7, pp.~219--242. 

\bibitem{AAW}
A.K.~Alves~de~Medeiros, W.M.P.~van~der~Aalst, and A.J.M.M.~Weijters (2008):
\newblock \emph{Quantifying Process Equivalence Based on Observed Behavior}.
\newblock {\sl Data \& Knowledge Engineering} 64, pp.~55--74.

\bibitem{BPW} 
M.~Backes, B.~Pfitzmann, and M.~Waidner (2007):
\newblock \emph{The Reactive Simulatability (RSIM) Framework for Asynchronous Systems}.
\newblock {\sl Information and Computation} 205, pp.~1685--1720.

\bibitem{HPA}
J.A.~Bergstra, A.~Ponse, and S.A.~Smolka, Eds. (2001):
\newblock \emph{Handbook of Process Algebra}.
\newblock {\sl Elsevier}.

\bibitem{Ber07}
M.~Bernardo (2007):
\newblock \emph{Non-Bisimulation-Based Markovian Behavioral Equivalences}.
\newblock {\sl Journal of Logic and Algebraic Programming} 72, pp.~3--49.

\bibitem{Buc} P.~Buchholz (1994):
\newblock \emph{Exact and Ordinary Lumpability in Finite Markov Chains}.
\newblock {\sl Journal of Applied Probability} 31, pp.~59--75.

\bibitem{dRT}
M.~de Rougemont and M.~Tracol (2009):
\newblock \emph{Static Analysis for Probabilistic Processes}.
\newblock {\sl Int.~Symp.~on Logic in Computer Science (LICS'09), IEEE-CS}, pp.~299--308.

\bibitem{DGJP}
J.~Desharnais, V.~Gupta, R.~Jagadeesan, and P.~Panangaden (2004):
\newblock \emph{Metrics for Labelled Markov Processes}.
\newblock {\sl Theoretical Computer Science} 318, pp.~323--354.

\bibitem{DLT}
J.~Desharnais, F.~Laviolette, and M.~Tracol (2008): 
\newblock \emph{Approximate Analysis of Probabilistic Processes: Logic, Simulation and Games}. 
\newblock {\sl Int.~Conf.~on Quantitative Evaluation of Systems (QEST'08), IEEE-CS}, pp.~264--273.

\bibitem{DHW} A.~Di~Pierro, C.~Hankin, and H.~Wiklicky (2005):
\newblock \emph{Measuring the Confinement of Probabilistic Systems}.
\newblock {\sl Theoretical Computer Science} 340, pp.~3--56.

\bibitem{DHW2} A.~Di~Pierro, C.~Hankin, and H.~Wiklicky (2008):
\newblock \emph{Quantifying Timing Leaks and Cost Optimisation}.
\newblock {\sl Conf.\ on Information and Comm.\ Security (ICICS'08), Springer LNCS} 5308, pp.~81--96.

\bibitem{LMMS} P.~Lincoln, J.C.~Mitchell, M.~Mitchell, and A.~Scedrov (1999):
\newblock \emph{Probabilistic Polynomial-time Equivalence and Security Analysis}.
\newblock {\sl World Congress on Formal Methods in the Development of Computing Systems (FM'99), Springer LNCS} 
1708, pp.~776--793.

\bibitem{FUMK} 
H.~Foster, S.~Uchitel, J.~Magee, and J.~Kramer (2003):
\newblock \emph{Model-based Verification of Web Service Compositions}.
\newblock {\sl Int.~Conf.~on Automated Software Engineering (ASE'03), IEEE-CS}, pp.~152--163.

\bibitem{ST}
R.~Segala and A.~Turrini (2007):
\newblock \emph{Approximated Computationally Bounded Simulation Relations for Probabilistic Automata}.
\newblock {\sl Computer Security Foundations Symposium (CSF'07), IEEE-CS}, pp.~140--156.

\bibitem{SWD} A.~Simpson, J.~Woodcock, and J.~Davies (1998):
\newblock \emph{Safety through Security}.
\newblock {\sl Workshop on Software Specification and Design (IWSSD'98), IEEE-CS} pp.~18--24.

\bibitem{Tzeng}
W.G.~Tzeng (1994):
\newblock \emph{A Polynomial-Time Algorithm for the Equivalence of Probabilistic Automata}.
\newblock {\sl SIAM Journal on Computing} 21, pp.~216--227.

\bibitem{vBW}
F.~van~Breugel and J.~Worrell (2005): 
\newblock \emph{A Behavioural Pseudometric for Probabilistic Transition Systems}. 
\newblock {\sl Theoretical Computer Science} 331, pp.~115--142.


\end{thebibliography}
